\documentclass[conference]{IEEEtran}
\IEEEoverridecommandlockouts
\usepackage{cite}
\usepackage{cuted}
\usepackage{makecell}
\usepackage{amsmath,amssymb,amsfonts}
\usepackage{algorithmic}
\usepackage{graphicx}
\usepackage{textcomp}
\usepackage{physics}
\usepackage{xcolor}
\usepackage{braket}
\def\BibTeX{{\rm B\kern-.05em{\sc i\kern-.025em b}\kern-.08em
    T\kern-.1667em\lower.7ex\hbox{E}\kern-.125emX}}

\newcommand{\INSTR}{\mathrm{INSTR}}

\begin{document}

\title{Instruction-Set Architecture for Programmable NV-Center Quantum Repeater Nodes
}

\author{
\IEEEauthorblockN{
Vinay Kumar\IEEEauthorrefmark{1}\IEEEauthorrefmark{2}, %\IEEEauthorrefmark{6}, %\thanks{\IEEEauthorrefmark{6}Email: vinay.kumar@phd.unipi.it},
Claudio Cicconetti\IEEEauthorrefmark{2},
Riccardo Bassoli\IEEEauthorrefmark{3}\IEEEauthorrefmark{4}\IEEEauthorrefmark{5},
Marco Conti\IEEEauthorrefmark{2},
and Andrea Passarella\IEEEauthorrefmark{2}
}
\vspace{0.7em}
\IEEEauthorblockA{\IEEEauthorrefmark{1}Department of Information Engineering, University of Pisa, Pisa, Italy\\
\IEEEauthorrefmark{2}Institute for Informatics and Telematics (IIT), National Research Council (CNR), Pisa, Italy\\
\IEEEauthorrefmark{3}Deutsche Telekom Chair of Communication Networks, Institute of Communication Technology,\\
Faculty of Electrical and Computer Engineering, Technische Universität Dresden, Dresden, Germany\\
\IEEEauthorrefmark{4}Centre for Tactile Internet with Human-in-the-Loop (CeTI), Cluster of Excellence, Dresden, Germany\\
\IEEEauthorrefmark{5}Quantum Communication Networks (QCNets) research group, Institute of Communication Technology,\\
Faculty of Electrical and Computer Engineering, Technische Universität Dresden, Dresden, Germany
}
}

\maketitle

\begin{abstract}
Programmability is increasingly central in emerging quantum network software stacks, yet the node-internal controller-to-hardware interface for quantum repeater devices remains under-specified. We introduce the idea of an instruction-set architecture (ISA) for controller-driven programmability of nitrogen-vacancy (NV) center quantum repeater nodes. Each node consists of an optically interfaced electron spin acting as a data qubit and a long-lived nuclear-spin register acting as a control program. We formalize two modes of programmability: (i) deterministic register control, where the nuclear register is initialized in a basis state to select a specific operation on the data qubit; and (ii) coherent register control, where the register is prepared in superposition, enabling coherent combinations of operations beyond classical programmability. Network protocols are expressed as controller-issued instruction vectors, which we illustrate through a compact realization of the BBPSSW purification protocol. We further show that coherent register control enables interferometric diagnostics such as fidelity witnessing and calibration, providing tools unavailable in classical programmability. Finally, we discuss scalability to multi-electron and multi-nuclear spin architectures and connection to Linear combination of unitaries (LCU) and Kraus formulation.
\end{abstract}


\section{Introduction}

The vision of a large-scale quantum internet rests on the ability to distribute, store, and process entanglement across many nodes. This vision has motivated a wide range of research on entanglement distribution, purification, quantum memories, and network protocols for routing, forwarding, and scheduling \cite{vkinternet25, rohdeinternet25, vkthesis}. At the heart of these protocols is the execution of local quantum operations within repeater nodes. While the action of such operations is algebraically straightforward, their physical realization on platforms such as ion traps, NV-centers, neutral atoms, or superconducting qubits is far more intricate and complex.  

Among the candidate physical systems, NV-centers in diamond have emerged as a leading platform for quantum repeaters, owing to their combination of optically addressable electron spins and long-lived nuclear-spin registers while operating at room temperature \cite{rozpedekNVrepeaterexp19, nemotoNVquantumentwork16, jingNVrepeater22}. These hybrid systems offer both optical interfaces for entanglement distribution and nuclear memories for storage, making them natural building blocks for a quantum internet.  

In classical networks, scalability and flexibility have been enabled by programmability. The paradigm of software-defined networking (SDN) introduced centralized control and dynamic reconfiguration \cite{mckeownopenflow08, kreutzSDN14}, while protocol-independent data planes such as P4 \cite{Bosshartp414} allowed programmable packet processing. These advances established the separation between control and data planes, enabling networks to adapt to diverse protocols and applications. Recent work has introduced operating-system style architectures for executing quantum network applications in hardware-independent high-level software, such as QNodeOS~\cite{delleOS25}. These efforts focus on application execution, scheduling, and process management across nodes. In contrast, this work addresses the node-internal instruction boundary: we formalize a compact ISA for NV repeater nodes, where controller-issued instructions map directly to electron–nuclear primitives and optional coherent diagnostic semantics. Inspired by this layered view, we ask: \emph{how can a controller-level instruction interface be defined inside a quantum repeater node, so that higher-layer protocols can flexibly direct the local quantum operations?}  

To address this question, this work makes the following contributions:
\begin{enumerate}
    \item \textbf{Instruction-set architecture (ISA).} We propose the idea of an ISA for NV-center quantum repeater nodes, exposing the set of available local operations to a controller and thereby enabling controller-driven programmability at the hardware level.
    \item \textbf{Two modes of programmability.} 
    a) \emph{Deterministic register control:} The nuclear-spin program register is initialized in a basis state to deterministically select a specific local operation, providing classical programmability analogous to SDN flow rules. b) \emph{Coherent register control:} The program register is prepared in superposition, enabling coherent combinations of operations and interference-based effects that go beyond classical programmability.
    \item \textbf{Diagnostics and fidelity witnessing.} We show that coherent programmability enables interference-based diagnostics, such as fidelity witnessing and calibration between unitaries, which are inaccessible under classical programmability.  
\end{enumerate}

The remainder of the paper is organized as follows. Section~\ref{sec:background} presents background and related work. Section~\ref{sec:isa} introduces the ISA-style abstraction for NV-center quantum repeater nodes. Section~\ref{sec:discussion} generalizes to multi-electron/multi-nuclear spin setups and discusses the reformulation of programmability in terms of LCU and Kraus operators. Section~\ref{sec:conclusion} concludes the paper.

\section{Background and Related Work}\label{sec:background}

NV-centers in diamond have emerged as a powerful solid-state platform for quantum information
processing and quantum networking. Over the past two decades, extensive research has established their capabilities for
entanglement distribution, quantum memory, and multi-qubit control. Bernien et al. \cite{bernienheralded13} achieved heralded entanglement between NV-centers separated by three meters by simultaneous measurement of two photons, each of which is entangled with an NV-center,
and within the same group, Hensen et al. \cite{hensenloophole15} reported a loophole-free violation of Bell’s inequality with electron spins
separated by 1.3 km. Kalb et al. \cite{kalbdistillation17} further demonstrated entanglement distillation between solid-state quantum
network nodes.

Beyond experiments, theoretical proposals have outlined how NV-centers can serve as quantum repeaters combining
entanglement generation, storage, purification, and swapping \cite{childressfault05, nemotoNVquantumentwork16}. These architectures envision entanglement swapping
via Bell-state measurements on pairs of qubits entangled with neighboring nodes.
These results provide the foundation for viewing NV-centers as building blocks of large-scale entanglement distribution
systems.

Complementary to experiments distributing entanglement, a broad range of studies have characterized the capabilities achievable within a single NV-center, forming the relevant operational regime for this work. This includes \emph{optical initialization and readout}\cite{jelezkoreview06}, \emph{single-qubit control}\cite{dobrovitskicontrol13}, \emph{electron-nuclear coupling}\cite{childresscoherent06}, \emph{quantum memory}\cite{maurermemory12}, \emph{error correction and repetitive readout}\cite{neumannreadout10, waldherrqec14}, and \emph{multi-qubit registers}\cite{bradleytenqubit19, abobeihmultiqubit18, taminiauuniversal14}. These studies suggest that a single NV-center can serve as a programmable quantum processor with initialization, manipulation, and readout of the electron spin, conditional logic between nuclear and electronic spins, and limited entanglement operations within the local register.
However, such systems do not inherently enable entanglement swapping between remote nodes without photonic interfaces.

Architecturally, field-programmable spin array designs aim to realize reconfigurable spin-based processors by tuning local parameters across NV arrays \cite{wangprogrammable23}. At the network level, routing and entanglement distribution have been explored via centralized, software-defined-network-style control planes with global link-state knowledge \cite{pantrouting19} and via fully distributed, decentralized protocols \cite{chakrabortyrouting19}. Beyond these approaches, quantum-native control architectures have also been proposed, where the control plane itself is placed in superposition. For instance, \cite{caleffiquantumaddressing25} introduces an entanglement-defined controller (EDC) that manages a quantum control plane with superposed network addresses, enabling quantum-native routing. While their architecture introduces superposition at the network layer, our framework keeps the controller classical and introduces superposition within the node, that is, in the nuclear-spin register, to achieve programmability, diagnostics, and LCU-type formulations.

To our knowledge, a controller-driven instruction-set abstraction at the node level, where a classical network controller selects from a physical set of implementable operations by preparing a nuclear-spin register, has not been formalized in one coherent framework. This is the gap we address in this work.

%Each programmable NV-center node functions as a quantum repeater with local programmability: it performs entanglement swapping and controlled operations on stored qubits under a classical instruction-set interface. While quantum routers \cite{hubermanrouter20} extend this model to include address-based forwarding across multiple destinations, our focus is on the node-level programmability that enables such routers to be constructed from locally calibrated, software-defined repeater elements.

\begin{figure}[tb]
    \includegraphics[width=\columnwidth ]{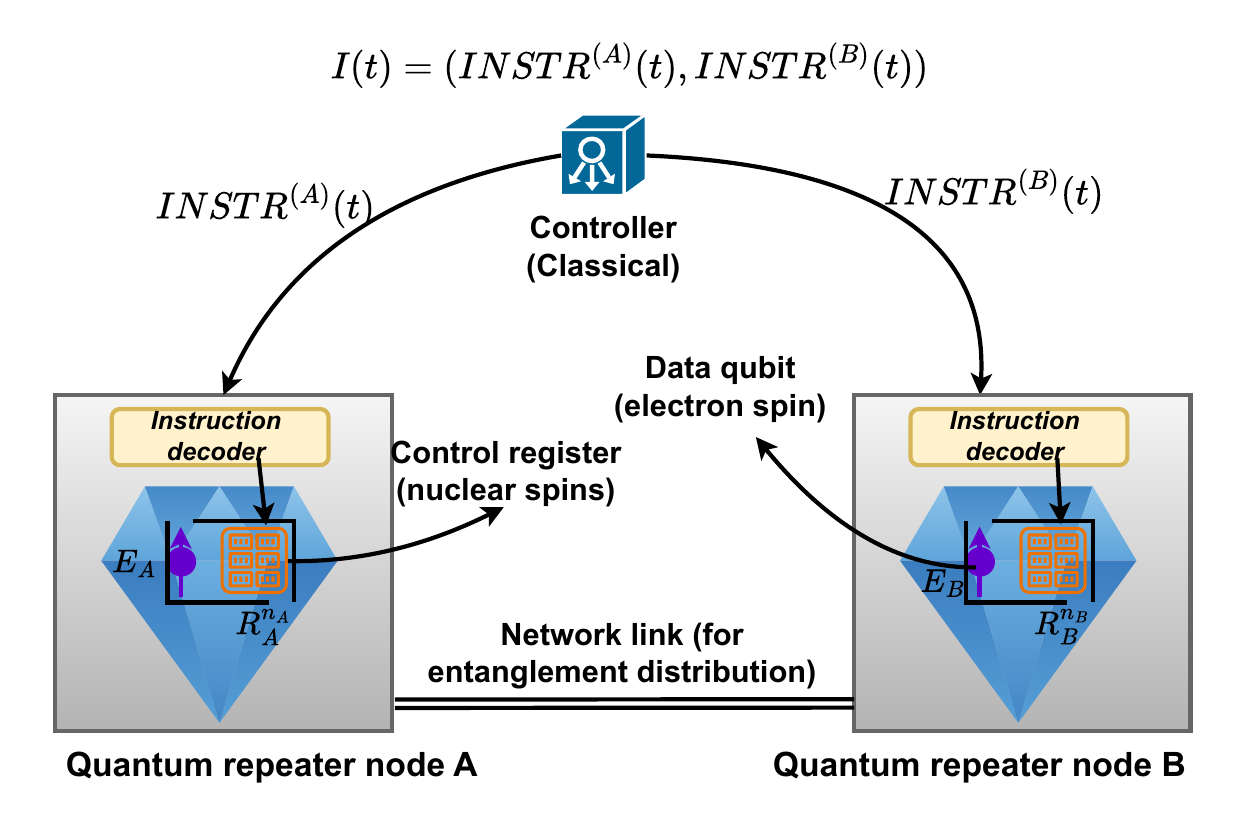}
    \caption{Two node example. A classical controller broadcasts the instruction vector \(\mathbf{I}(t)\). Each node decodes it into nuclear register preparation and microwave (MW) or radio-frequency (RF) pulse sequences that realize the selected electron spin operation.}
    \label{fig:isa}
    %\vspace{-0.5cm}
\end{figure}

\section{Instruction-set architecture (ISA) for NV-Center Nodes}\label{sec:isa}

Consider an NV-center node with one electron spin $E_0$ and $n$ nuclear spins $N_1, \dots, N_n$, which together form a control register. 
Then the Hilbert space is given by
\begin{equation}
    \mathcal{H} = \mathcal{H}_{E_{0}} \otimes \left( \bigotimes_{i=1}^{n} \mathcal{H}_{N_i} \right),
\end{equation}
with $\dim (\mathcal{H}) = 2^{1+n}$. Here, $0$ signifies the index of electron spin used.

A quantum network comprises $M$ such nodes, each controlled by a centralized classical controller, as shown in Fig.~\ref{fig:isa} for a two-node example. 
At any time-step $t$, the controller broadcasts an instruction vector
\begin{equation}
\mathbf{I}(t) = \big(\INSTR^{(1)}(t), \dots, \INSTR^{(M)}(t)\big),
\end{equation}
where each element $\INSTR^{(m)}(t)$ specifies the operation to be executed by node $m$. The controller-driven execution model implicitly assumes a time-slotted
schedule: each instruction round \(t\) corresponds to a network-wide control slot during which all nodes complete the operations specified by \(\mathbf{I}(t)\) before proceeding to the next round. While asynchronous
execution is theoretically possible, synchronization ensures that all spin operations complete within the coherence time of the most short-lived qubit involved in the protocol, thereby preserving end-to-end or protocol fidelity. This time-slotted abstraction aligns with the architecture advocated in \cite{Beauchamparchitecture25}, where slotted control
is shown to be beneficial for near-term quantum networks.
Each node contains a \emph{local decoder} that interprets its incoming instruction and configures both the nuclear register and the control pulse sequence required to enact the instruction.

\paragraph*{Instruction format}
Let $n$ denote the number of nuclear control qubits assigned to electron spin qubit $E_0$, so the register exposes $2^n$ addresses. Each instruction sent to an electron spin has the structure
\begin{equation}
\begin{aligned}
\INSTR^{(0)} = \big(
&\textsf{OPCODE},\;
 \textsf{PARAMS},\\
&\textsf{PATTERN}\subseteq\{0,\dots,2^n\!-\!1\},\;
 \textsf{MODE}
\big),
\end{aligned}
\label{eq:instr_def}
\end{equation} where:
\begin{itemize}
\item \textsf{OPCODE} selects a primitive gate, e.g.\ $X$, $Z$, $H$, $R_y(\theta)$, $\text{CNOT}(E_0\!\to\!N_A)$, or $\text{MEASURE}$;
\item \textsf{PARAMS} carries any continuous parameters such as rotation angles or phases, and addressing information for multi-qubit gates, such as the identifiers of control and target qubits (\texttt{control = $E_0$}, \texttt{target = $N_A$});
\item \textsf{PATTERN} identifies the subset of nuclear configurations that enable the operation;
\item \textsf{MODE}\,$\in\{\textsf{deterministic},\textsf{coherent}\}$ specifies the mode of programmability of nuclear register.
\end{itemize}

When \textsf{MODE} is \textsf{deterministic}, the local decoder initializes the nuclear register into one configuration from the set defined by \textsf{PATTERN}, producing a deterministic operation. 
When \textsf{MODE} is \textsf{coherent}, the decoder prepares a coherent superposition over the nuclear register configurations, allowing the corresponding electron spin operations to occur in superposition within the same execution cycle.

Physically, these logical instructions are realized through local microwave (MW) and radio-frequency (RF) control fields. 
Each node includes classical electronics that generate the MW pulses required for manipulating the electron spin and the RF pulses used to drive nuclear-spin transitions. 
The decoder configures these pulse generators according to $\INSTR^{(0)}(t)$ so that the required initialization and controlled electron operation are performed. 
In this way, the instruction-set abstraction encapsulates both the logical operation and its hardware realization through MW/RF control at each node.

\paragraph*{Execution semantics}
An instruction acting in \textsf{coherent} mode realizes a conditional unitary of the form \cite{nielsenarray97}
\begin{equation}
U^{(0)} = \sum_{a_i\in \textsf{PATTERN}} |a_i\rangle\!\langle a_i|_N \otimes U_{E_0}(a_i),
\end{equation}
where $U_{E_0}(a_i)$ corresponds to the primitive selected by \textsf{OPCODE}$,$ parameterized by \textsf{PARAMS}. In \textsf{deterministic} mode the same instruction specializes to a single selected branch \(U^{(0)} = \ket{a}\!\bra{a}_N \otimes U_{E_0}(a)\) for some \(a \in \textsf{PATTERN}\).

The global network evolution at time-step $t$ is then
\begin{equation}
U_{\text{network}}(t) = \bigotimes_{m=1}^M U^{(m)}_{\INSTR^{(m)}(t)},
\end{equation}
and a complete protocol of $T$ instruction rounds is represented as
\begin{equation}
U_{\text{protocol}} = U_{\text{network}}(T)\cdots U_{\text{network}}(1).
\end{equation}

\subsection{Deterministic Register Control}\label{ssec:deterministiccontrol}

In the deterministic mode defined by the ISA, the nuclear-spin register serves as a classical control program while the electron spin functions as the data qubit; this reflects the operational regime of current NV-center experiments, where the register is prepared in definite basis states and used as classical control memory. 
The controller initializes the nuclear register into a specific configuration selected by the instruction’s \textsf{PATTERN} field (one of the allowed basis states), and the corresponding electron spin operation is applied according to the \textsf{OPCODE} and \textsf{PARAMS}. 
This realizes classical programmability, where each nuclear configuration \emph{deterministically} selects a single operation on the electron spin.

The required conditional gates, such as controlled rotations or CNOT operations, are implemented through the same MW and RF pulse sequences described in Section~\ref{sec:isa}. 
To illustrate, consider a node consisting of two nuclear spins ($N_1$, $N_2$) and one electron spin ($E_0$).  
Before instruction execution, the joint state is initialized as
\begin{equation}
\ket{\Psi_{\text{initial}}} =
\ket{0}_{N_1} \ket{0}_{N_2} \otimes
\bigl( \alpha \ket{0}_{E_0} + \beta \ket{1}_{E_0} \bigr),
\label{eq:deterministic_initial}
\end{equation}
where $\ket{0}_{N_1} \ket{0}_{N_2}$ denotes one of the four basis configurations of the nuclear register.  
The action taken by the node depends deterministically on the nuclear configuration. \noindent For a two-nuclear-spin register $(N_1,N_2)$, the four basis states select different operations on the electron spin:
\begin{enumerate}
    \item $00$: idle ($I$),
    \item $01$: bit flip ($X$),
    \item $10$: rotation $R_y(\theta)$,
    \item $11$: entangling operation (e.g., $\text{CNOT}_{E_0\rightarrow N_A}$ with the electron spin as control and an ancillary nuclear spin $N_A$ as target).
\end{enumerate}

In this mode, each nuclear configuration acts as a classical program entry that deterministically selects a single operation.

\begin{figure}[tb]
    \includegraphics[width=\columnwidth]{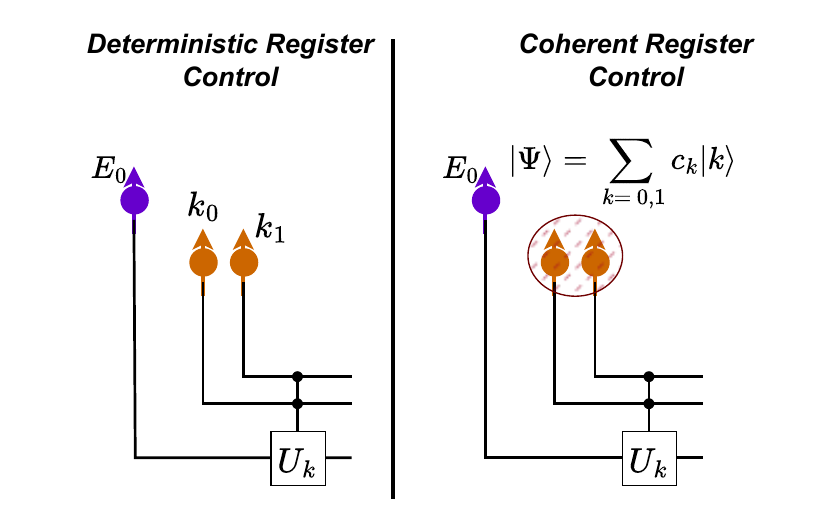}
    \caption{Programmability modes. Deterministic mode initializes the nuclear register in a basis state that selects one operation. Coherent mode prepares and reads the register in a rotated basis to enact a linear combination of operations on the electron spin.}
    \label{fig:deterministic_coherent_control}
    %\vspace{-0.5cm}
\end{figure}

\subsection{Coherent Register Control}\label{ssec:coherentcontrol}

Building upon the deterministic case in Section~\ref{ssec:deterministiccontrol}, we now extend to \emph{coherent register control}, where the nuclear spins are prepared in superposition rather than fixed classical configurations. Deterministic register control is the default execution mode for network protocols because it maximizes repetition rate and avoids postselection overhead (discarded outcomes), whereas coherent register control is invoked selectively for diagnostic and calibration tasks. In this mode, the nuclear register acts as a quantum address register, coherently selecting among multiple electron-spin side unitaries within a single execution cycle.

\paragraph*{Superposed nuclear register}
Consider again two nuclear spins $N_1$ and $N_2$ prepared independently in arbitrary superpositions:
\[
|\Psi_{N_1 N_2}\rangle = 
\big(\alpha_0 |0\rangle + \alpha_1 |1\rangle\big)_{N_1}
\otimes
\big(\beta_0 |0\rangle + \beta_1 |1\rangle\big)_{N_2}, \ \text{or},
\]
\begin{equation}\label{eq:N1N2_superposition}
|\Psi_{N_1 N_2}\rangle
= \alpha_0 \beta_0 |00\rangle
+ \alpha_0 \beta_1 |01\rangle
+ \alpha_1 \beta_0 |10\rangle
+ \alpha_1 \beta_1 |11\rangle,
\end{equation}

where the amplitudes $\alpha_i \beta_j$, $\forall \ i, \ j \in \{0, 1\}$ represent the probability amplitudes of selecting nuclear bits in basis $|ij\rangle$. Now, each computational basis configuration $\ket{N_1 N_2}$ corresponds to a deterministic operation from Section~\ref{ssec:deterministiccontrol}. We represent these collectively by a controlled operation \cite{nielsenarray97}
\begin{equation}\label{eq:U_router}
U_{\mathrm{repeater}} = 
\sum_{i,j \in \{0,1\}} |ij\rangle\!\langle ij| \otimes U_{ij},
\end{equation}
where
$U_{00}=I, \quad
U_{01}=X, \quad
U_{10}=R_y(\theta), \quad
U_{11}=\text{CNOT}_{E_{0}\to N_A}.$
Let the electron spin be initialized as $|\psi_{E_0}\rangle = \gamma_0 |0\rangle + \gamma_1 |1\rangle$ so that the joint initial state is $|\Psi_{\mathrm{initial}}\rangle
= |\Psi_{N_1 N_2}\rangle \otimes |\psi_{E_0}\rangle$.

Applying $U_{\mathrm{repeater}}$ to it gives
\begin{align}\label{eq:entangled_router}
\begin{aligned}
|\Psi_{\text{after}}\rangle = {} & \alpha_0 \beta_0 |00\rangle \otimes I_E |\psi_{E_0}\rangle \\
&+ \alpha_0 \beta_1 |01\rangle \otimes X |\psi_{E_0}\rangle \\
&+ \alpha_1 \beta_0 |10\rangle \otimes R_y(\theta) |\psi_{E_0}\rangle \\
&+ \alpha_1 \beta_1 |11\rangle \otimes \text{CNOT}_{{E_0} \to N_A} |\psi_{E_0}\rangle.
\end{aligned}
\end{align}

Equation~\eqref{eq:entangled_router} shows that the repeater is now placed in a coherent superposition of performing different electron-side operations, weighted by the amplitudes of the nuclear register. This is the key distinction from deterministic control, where only one branch is active per run.

\paragraph*{Generalized formulation}
The general controlled operation is (Eq.~\eqref{eq:U_router})
\begin{equation}\label{eq:U_r_generalised}
U_{\mathrm{repeater}} = \sum_{k=0}^{K-1} \ketbra{k} \otimes U_k,
\end{equation}
acting on a register prepared in superposition
$
\ket{\psi_N} = \sum_{k=0}^{K-1} c_k \ket{k},
\qquad \sum_k |c_k|^2 = 1.
$
The post-operation state is then
\begin{equation}\label{eq:qaddr_general}
U_{\mathrm{repeater}}\big(\ket{\psi_N}\otimes\ket{\psi_{E_0}}\big)
= \sum_{k=0}^{K-1} c_k\, \ket{k}\otimes U_k\ket{\psi_{E_0}}.
\end{equation}
%Measuring the nuclear register in the computational basis collapses the system to one branch $U_k$, reproducing classical programmability. The advantage of coherent control emerges only when the nuclear register is measured in a rotated basis, producing interference between the electron-side operations.

\paragraph*{Measurement in a rotated basis}
Let the nuclear register be projected onto a general readout state
$
\ket{\phi} = \sum_{k=0}^{K-1} d_k \ket{k},
\qquad \sum_k |d_k|^2 = 1.
$
Projecting Eq.~\eqref{eq:qaddr_general} gives the (unnormalized) electron state
\begin{equation}\label{eq:Kraus-operator}
(\bra{\phi}\otimes I)\,U_{\mathrm{repeater}}(\ket{\psi_N}\otimes\ket{\psi_{E_0}})
= \Big(\sum_{k=0}^{K-1} d_k^{\!*}c_k\,U_k\Big)\ket{\psi_{E_0}}.
\end{equation}
with the corresponding outcome probability is
\begin{equation}
p_\phi = \big\|(\sum_k d_k^{\!*}c_k\,U_k)\ket{\psi_{E_0}}\big\|^2.
\end{equation}

\paragraph*{Example: Two-branch interference}
Consider a single nuclear qubit with unitaries $\{U_0,U_1\}$ and program state $\ket{\psi_N}=(\ket{0}+\ket{1})/\sqrt{2}$. After applying $U_{\mathrm{repeater}}$,
\begin{equation}\label{eq:two-branch-after}
\begin{aligned}
\ket{\Psi_{\mathrm{after}}}
&= U_{\mathrm{repeater}}\big(\ket{\psi_N}\otimes\ket{\psi_{E_0}}\big) \\
&= \tfrac{1}{\sqrt{2}}\Big( \ket{0}\otimes U_0\ket{\psi_{E_0}}
   + \ket{1}\otimes U_1\ket{\psi_{E_0}} \Big).
\end{aligned}
\end{equation}
Measuring the nuclear spin in the $X$ basis,
\(
\ket{\phi_\pm}=(\ket{0}\pm\ket{1})/\sqrt{2},
\)
yields the unnormalized electron states
\begin{equation}
 \Rightarrow \ket{\psi_{E_0}^{(\pm)}} = (\bra{\phi_\pm}\otimes I)\ket{\Psi_{\mathrm{after}}} = \frac{1}{2}\Big(U_0 \;\pm\; U_1\Big) \ket{\psi_{E_0}},
\end{equation}
with corresponding probabilities as

\begin{equation}
    p_\pm
= \left\langle \psi_{E_0} \left| \tfrac{1}{2}(U_0^\dagger \pm U_1^\dagger)\, 
                      \tfrac{1}{2}(U_0 \pm U_1) \right| \psi_{E_0} \right\rangle \text{or},
\end{equation}

\begin{equation}\label{eq:p_pm_basic}
\Rightarrow p_\pm = \tfrac{1}{2}\big(1 \pm \mathrm{Re}\langle\psi_{E_0}|U_0^\dagger U_1|\psi_{E_0}\rangle\big).
\end{equation}
The interference term $\mathrm{Re}\langle\psi_{E_0}|U_0^\dagger U_1|\psi_{E_0}\rangle$ encodes the overlap between the two implemented unitaries, that is, the information inaccessible in deterministic control.

\paragraph*{Phase scanning and fidelity witnessing}
If the nuclear program qubit is prepared with a relative phase,
\(
\ket{\psi_N(\varphi)}=(\ket{0}+e^{i\varphi}\ket{1})/\sqrt{2},
\)
the probabilities generalize to
\begin{equation}\label{eq:p_pm_phi_final}
p_\pm(\varphi)
= \tfrac{1}{2}\big(1 \pm \mathrm{Re}[e^{i\varphi}\langle\psi_{E_0}|U_0^\dagger U_1|\psi_{E_0}\rangle]\big).
\end{equation}
By sweeping $\varphi$, both the real and imaginary parts of $\langle\psi_{E_0}|U_0^\dagger U_1|\psi_{E_0}\rangle$ can be extracted experimentally. Using two phase settings $\varphi_1=0$ and $\varphi_2=-\pi/2$, one obtains
\begin{equation}
\mathrm{Re}\,a = 2p_+^{(1)} - 1,\quad
\mathrm{Im}\,a = 2p_+^{(2)} - 1,\quad
a = \langle\psi_{E_0}|U_0^\dagger U_1|\psi_{E_0}\rangle.
\end{equation}

Hence, the state:
\begin{equation}\label{eq:amplitude_calculated}
    \Rightarrow \bra{\psi_{E_0}} U_0^{\dagger} U_1 \ket{\psi_{E_0}} = (2 p_+^{(1)} - 1) + i\, (2 p_+^{(2)} - 1).
\end{equation}
\paragraph*{Fidelity between the two applied unitaries}
Let's say we have two implemented unitaries $U_0$ and $U_1$ acting on the same electron input $\ket{\psi_{E_0}}$, and we want to quantify how close their outputs are. The relevant \emph{state fidelity} is as follows:
\begin{equation}
F_{\mathrm{state}}(\psi_{E_0};U_0,U_1) =
\big|\bra{\psi_{E_0}}U_0^\dagger U_1\ket{\psi_{E_0}}\big|^2,
\label{eq:Fstate-def}
\end{equation}
which is straightforward from the eq~\eqref{eq:amplitude_calculated}:
\begin{equation}
    F_{\mathrm{state}}(\psi_{E_0};U_0,U_1) = (2 p_+^{(1)} - 1)^2 + (2 p_+^{(2)} - 1)^2.
\end{equation}

\paragraph*{Interpretation}
Coherent register control thus enables the NV-center node to act as an \emph{interferometric fidelity witness}, comparing the action of two or more operations on a chosen input state without full process tomography. The nuclear register serves as a coherent selector that allows interference between different branches of control, providing in situ calibration and diagnostic capability not possible under deterministic, classical programmability.
\begin{center}
\fbox{%
\parbox{0.99\columnwidth}{%
\footnotesize
\textbf{Note:}
By this approach, the repeater does not give direct access to the
\emph{global} gate fidelity
\(F(U_0,U_1)=|\mathrm{Tr}(U_0^\dagger U_1)|^2/d^2\),
which would require full process tomography. Instead, by preparing a chosen
input state \(\ket{\psi_{E_0}}\), the repeater yields the state-dependent overlap
\(\langle\psi_{E_0}|U_0^\dagger U_1|\psi_{E_0}\rangle\). This serves as a
\emph{fidelity witness}: it certifies the closeness of \(U_0\) and \(U_1\) on the probed
state, providing partial diagnostic information without reconstructing the
complete channels.
}}
\end{center}
This capability is also useful for \emph{in situ calibration}: the applied pulse or pulse sequence can be tuned so that the implemented operation better matches the intended unitary. In addition, having a coherent repeater enables situations where the node can coherently evaluate or compare multiple unitaries before committing to one, something that is impossible under classical register control. In the classical case, once the nuclear register is prepared in a definite basis state, the corresponding unitary is fixed, and no further adjustment or comparison is possible.
\subsection{Controller-Driven Network Protocol Example}\label{ssec:controllerdrivenexample}

One of the most common protocols in quantum networks is entanglement purification. In this section, we showcase the typical implementation of the ISA by using the example of the BBPSSW purification protocol~\cite{Bennettpurification96}.

\begin{figure}[tb]
    \includegraphics[width=\columnwidth]{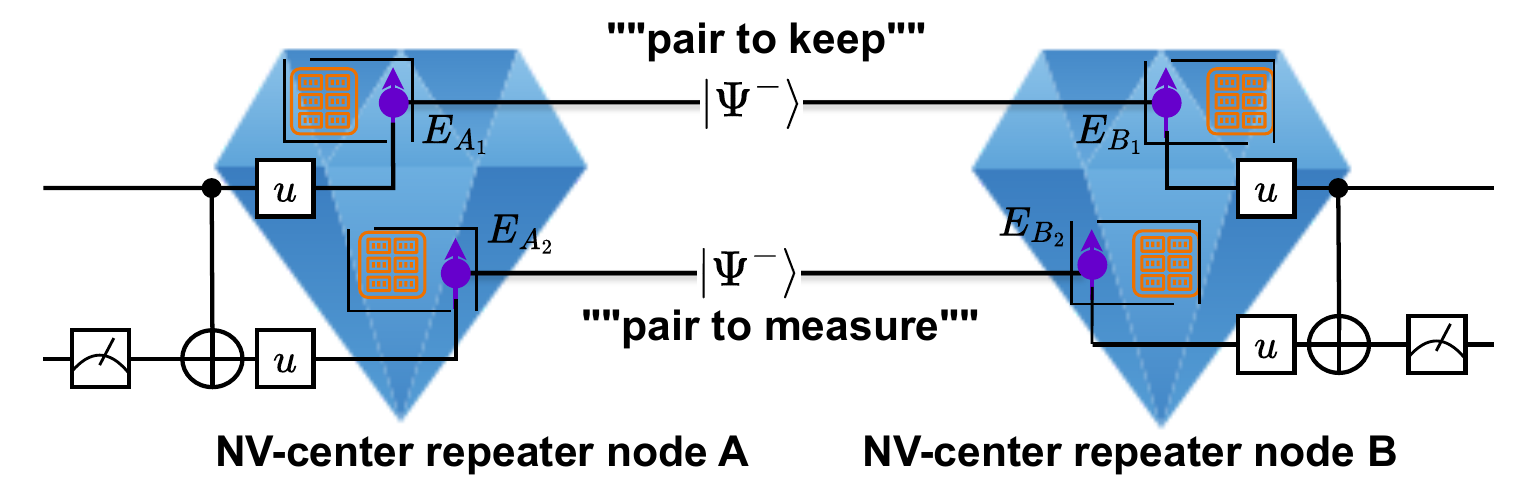}
    \caption{An example: BBPSSW purification protocol in NV-center nodes.}
    \label{fig:isa_purification}
    %\vspace{-0.5cm}
\end{figure}

Consider two NV-center nodes, $A$ and $B$, that each hold two entangled electron-spin pairs shared between them through a noisy quantum channel as shown in Fig.~\ref{fig:isa_purification}:
\begin{itemize}
    \item Pair~1: $(E_{A1}, E_{B1})$: the pair to keep.
    \item Pair~2: $(E_{A2}, E_{B2})$: the pair to measure.
\end{itemize}

The controller sends out a vector of structured instructions to both nodes to execute the Bennett \emph{et al.} purification protocol. Each instruction to electron-spin $E_e$ at node $m$ follows the format
\[
\INSTR^{(m, e)} = \big(\textsf{OPCODE},\; \textsf{PARAMS},\; \textsf{PATTERN},\; \textsf{MODE}\big),
\]
where in this case \textsf{MODE}=\textsf{deterministic}. The local decoder at each node interprets the received instruction to configure the relevant electron and nuclear spins and to generate the corresponding MW/RF pulse sequence as summarized in table~\ref{tab:BBPSSW_Uk}. 

\noindent\textbf{Remark.}
The BBPSSW example uses a minimal sufficient subset of primitives; additional protocol suites may require further opcodes (for example, explicit memory moves or multi-qubit measurements), while the ISA structure in Eq.~\eqref{eq:instr_def} remains unchanged.

%The protocol proceeds as follows:
\begin{table}[t]
\centering
\caption{Instruction set \(U_k\) sufficient to implement the BBPSSW protocol.}
\label{tab:BBPSSW_Uk}
\renewcommand{\arraystretch}{1.1}
\setlength{\tabcolsep}{4pt}
\footnotesize
\begin{tabular}{c c p{0.55\columnwidth}}
\hline
\makecell{\textbf{Index} \\ (\textsf{PATTERN})} &
\makecell{\textbf{Operation} \\ (\textsf{OPCODE})} &
\textbf{Description} \\
\hline
000 & \(I\)                & Idle or no operation \\
001 & \(X\)                & Bit flip (Pauli \(X\)) \\
010 & \(Y\)                & Bit and phase flip (Pauli \(Y\)) \\
011 & \(Z\)                & Phase flip (Pauli \(Z\)) \\
100 & \(\mathrm{CNOT}\)    & Two-qubit entangling gate \\
101 & \(\mathrm{MEASURE}\) & Projective measurement in the computational basis \\
\hline
\end{tabular}
\end{table}
The controller issues a sequence of instructions implementing the BBPSSW protocol:

\begin{enumerate}
\item Bilateral twirling: both nodes apply the same random Pauli rotation to each pair.
\item Bilateral CNOT: the “keep” qubit controls the “measure” qubit at each node.
\item Local measurement: the target qubits are measured and classical outcomes exchanged.
\item Parity check: pairs are kept only when outcomes agree.
\end{enumerate}
The required instruction set is therefore
\[
\{\textsf{PAULI},\, \textsf{CNOT},\, \textsf{MEASURE}\}.
\]

Let the corresponding local quantum operation at node $m$ be denoted by $\mathcal{E}^{(m)}_{\textsf{OPCODE}}$. For $\textsf{PAULI}$ and $\textsf{CNOT}$, the operation is unitary:
\begin{equation}
\mathcal{E}^{(m)}_{\textsf{OPCODE}}(\rho)
= U^{(m)}_{\textsf{OPCODE}}\rho\,U^{(m)\dagger}_{\textsf{OPCODE}},
\label{eq:Emk_def_new}
\end{equation}
where $U^{(m)}_{\textsf{OPCODE}} \in \{I,X,Y,Z,\mathrm{CNOT}\}$ acts on the addressed qubit(s).

For $\textsf{MEASURE}$ in the computational basis,
\begin{equation}
\begin{aligned}
\mathcal{E}^{(m)}_{\textsf{MEASURE}}(\rho)
&= \sum_{b\in\{0,1\}}
   \big(\Pi_b\,\rho\,\Pi_b\big)
   \otimes \ket{b}\!\bra{b}_{C_m},\\
\Pi_0 &= \ket{0}\!\bra{0},\quad
\Pi_1 = \ket{1}\!\bra{1},
\end{aligned}
\end{equation}
optionally recording the classical outcome in a register $C_m$. Here, $\rho$ denotes the quantum state at node $m$, and $\Pi_0$ and $\Pi_1$ are the projectors onto the computational basis states $\ket{0}$ and $\ket{1}$, respectively. 
The sum over $b$ accounts for both possible outcomes of the measurement.
The term $\Pi_b\,\rho\,\Pi_b$ represents the \emph{state collapse} associated with outcome $b$, while the tensor product with $\ket{b}\!\bra{b}_{C_m}$ optionally records the classical result into a register $C_m$ associated with node $m$.
This record allows the measurement outcome to be used in later classical-control steps of a protocol (e.g., parity checks in BBPSSW).
\begin{center}
\fbox{%
\parbox{0.99\columnwidth}{%
\footnotesize
\textbf{Note:}
In Sections~\ref{ssec:deterministiccontrol} and~\ref{ssec:coherentcontrol}, all instructions corresponded to unitaries. Here, the \textsf{MEASURE} operation is non-unitary and irreversible. 
Accordingly, the model extends from a set of unitaries $U_k$ to a set of completely positive trace-preserving (CPTP) maps $\mathcal{E}_{\textsf{OPCODE}}$, encompassing both reversible and measurement-based actions under the same instruction framework.
}}
\end{center}

\section{Discussion and Outlook}\label{sec:discussion}
The formulation till now has been with only one electron spin per node. In Section~\ref{ssec:gen_espin}, we generalize to a node having multiple electron spins with a lightweight throughput model in Section~\ref{ssec:throughput_model}, followed by a discussion on connection to LCU and Kraus formulation in Section~\ref{ssec:kraus}. 
\subsection{Generalization to multiple electron spins per node}\label{ssec:gen_espin}

The ISA formulation so far has assumed a single electron spin denoted by $E_0$ and $n$ nuclear spins per node, as described in Sections~\ref{ssec:deterministiccontrol} and~\ref{ssec:coherentcontrol}. 
We now generalize to the case of $E_e$ electron spins per node, where each electron $E_e$ is paired with a local nuclear-spin register $\{N_{e,1}, \dots, N_{e,r}\}$ that provides its control. 
The ratio $r$ thus denotes the number of nuclear spins per electron.

The Hilbert space for a single electron–nuclear cluster is
\begin{equation}
\mathcal{H}_{\mathrm{cluster}}^{(e)} 
= \mathcal{H}_{E_e} \otimes \bigotimes_{i=1}^{r} \mathcal{H}_{N_{e,i}},
\end{equation}
where $\mathcal{H}_{E_e} \cong \mathbb{C}^2$ and $\mathcal{H}_{N_{e,i}} \cong \mathbb{C}^2$. 
The node-level Hilbert space is therefore
\begin{equation}
\mathcal{H}_{\mathrm{node}} = 
\bigotimes_{e=1}^{E} \mathcal{H}_{\mathrm{cluster}}^{(e)}, 
\qquad
\dim(\mathcal{H}_{\mathrm{node}}) = 2^{E(1+r)}.
\end{equation}

For each electron $E_e$, the control register spans a computational basis $\{\ket{k}_{N_e}\}_{k=0}^{2^r-1}$, defining the repeater operator corresponding to Eq.~\eqref{eq:U_r_generalised}:
\begin{equation}
U_{\mathrm{repeater}}^{(e)} = 
\sum_{k=0}^{2^r-1} \ketbra{k}_{N_e} \otimes U^{(e)}_k,
\end{equation}
where each $U^{(e)}_k$ is a local unitary (or joint electron–nuclear gate) determined by the controller-issued instruction $\INSTR^{(e)}$. 
Assuming independent control over each electron spin, the node-level evolution is
\begin{equation}
U_{\mathrm{node}} = 
\bigotimes_{e=1}^{E} U_{\mathrm{repeater}}^{(e)}.
\end{equation}

Each instruction $\INSTR^{(e)}$ takes the same structured form as defined in Eq.~\eqref{eq:instr_def}.

For a network with $M$ nodes, each containing $E_m$ electron spins and $r_m$ nuclear spins per electron, the global instruction vector at round $t$ is
\begin{equation}
\mathbf{I}(t) = 
\big\{\,\INSTR^{(m,e)}(t)\ \big|\ 
m=1,\dots,M,\ e=1,\dots,E_m\,\big\},
\end{equation}
and the corresponding evolution is
\begin{equation}
U_{\mathrm{network}}(t) = 
\bigotimes_{m=1}^{M} \ 
\bigotimes_{e=1}^{E_m} 
U^{(m,e)}_{\mathrm{r}}.
\end{equation}

Higher $E$ enables greater parallelism but increases the controller’s instruction-issuing overhead. 
Larger $r$ broadens the instruction set but lengthens nuclear re-initialization cycles, potentially limiting protocol repetition rates as evaluated in Section~\ref{ssec:throughput_model}.

Current diamond platforms have demonstrated multi-electron operation within a single crystal up to two NV electron spins, with coherent dipolar coupling and room-temperature entanglement \cite{dolde2013twoNV}.
On the memory side, a fully controlled ten-qubit (one electron + nine $^{13}$C nuclei) register has been operated as a processor with minute-scale quantum memory \cite{bradleytenqubit19}. Beyond that, isotopically engineered samples have been identified and individually addressed much larger neighborhoods of nuclear spins around a single NV; for example, \cite{abobeihfault22} reports control of dozens of $^{13}$C spins and use a subset to realize a fault-tolerant logical qubit.
Taken together, these results motivate our modeling choice of one electron spin with a multi-qubit nuclear register as the baseline node today, while the ISA generalization to $E_{e}>1$ electron spins per node should be read as a forward-looking architectural extension that becomes relevant as multi-electron spin NV modules mature \cite{taminiauuniversal14}.

\subsection{Performance Model: Throughput vs. Nuclear Re-initialization}\label{ssec:throughput_model}

In the time-slotted model (Sec.~\ref{sec:isa}), each instruction round $t$ first re-initializes the nuclear register, followed by local MW/RF control on the electron spin and nuclear register and optional readout. In the proposed ISA, the nuclear register acts as the program memory: its size $r$ determines both the instruction address space ($2^r$ possible configurations) and the time required to reinitialize the register between rounds. A simple node-level performance metric can be defined as the per-electron spin \emph{round throughput} $R$ (rounds/s), given by:
\begin{equation}
R \;\approx\; \frac{1}{t_{\mathrm{MW}} + t_{\mathrm{RF}} + t_{\mathrm{meas}} + t_{\mathrm{class}} + t_{\mathrm{reinit}}(r)},
\end{equation}
where $t_{\mathrm{MW}}$ and $t_{\mathrm{RF}}$ denote aggregated microwave and radio-frequency pulse times needed to execute the corresponding round (e.g., $X$, $Y$, and $\text{CNOT}$), $t_{\mathrm{meas}}$ captures any qubit readout invoked by the instruction, $t_{\mathrm{class}}$ is any classical-controller latency folded into the slot, and $t_{\mathrm{reinit}}(r) = \mathrm{\tau_{reset}} \ . \ r$ is the nuclear register re-initialization time for a register of size $r$ nuclear spins with per-nuclear spin reset time of $\mathrm{\tau_{reset}}$. So, $t_\mathrm{reinit}$ grows with $r$ (e.g., linearly if spins are reset sequentially), reflecting the cost of programmability.

If a node contains $E_e$ electron spins operating in parallel independently (Sec.~\ref{ssec:gen_espin}), then the idealized node round throughput simply scales as
\begin{equation}
R_{\mathrm{node}} \;\approx\; E_e \, R,
\end{equation}
subject to controller and crosstalk constraints between spins. For multi-round protocols such as BBPSSW purification (Sec.~\ref{ssec:controllerdrivenexample}), the time to complete one round is simply $1/R$.%; the time to reach a target fidelity then scales with the number of surviving rounds.

\paragraph*{Assumptions}
The model is intentionally minimal. We treat $t_{\mathrm{MW}}, t_{\mathrm{RF}}, t_{\mathrm{meas}}, t_{\mathrm{class}}$ as configuration-dependent constants for a given protocol step; the only variable we sweep is $t_{\mathrm{reinit}}(r)$ to highlight the programmability-rate trade-off. Coherent control adds a small fixed overhead (phase preparation and rotated-basis readout), which can be absorbed into $t_{\mathrm{MW}}{+}t_{\mathrm{RF}}{+}t_{\mathrm{meas}}$.

\begin{figure}[t]
  \centering
  \includegraphics[width=\columnwidth]{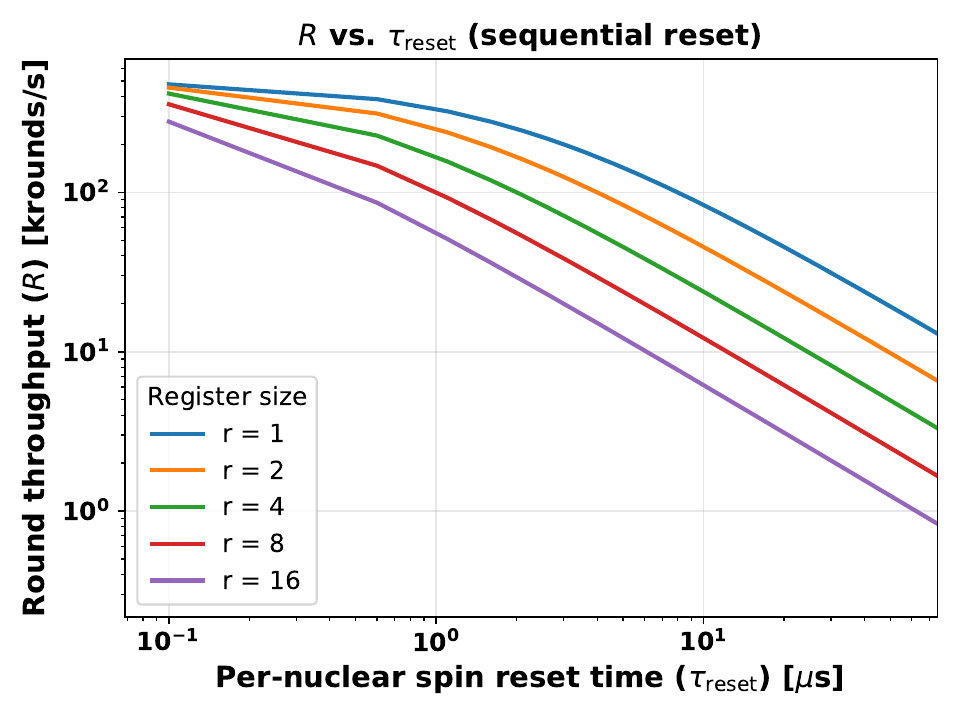}
  \caption{Per-electron round throughput $R$ vs Per-nuclear spin reset time $\tau_{\mathrm{reset}}$ for register sizes $r\in\{1,2,4,8, 16\}$ with fixed overheads ($t_{\mathrm{MW}}{+}t_{\mathrm{RF}}{+}t_{\mathrm{meas}}{+}t_{\mathrm{class}}$). Larger $r$ expands the ISA address space (more programmable operation choices) but increases re-initialization time.}
  \label{fig:throughput_vs_reinit}
  %\vspace{-0.5cm}
\end{figure}

Fig.~\ref{fig:throughput_vs_reinit} shows that as $\mathrm{\tau_{reset}}$ grows, throughput falls hyperbolically, with steeper degradation for larger $r$. Thus, beyond a register size that saturates the required instruction set, increasing $r$ yields diminishing returns: it broadens programmability while slowing repetition rate. The condition $2^r \ge p$ ensures that all $p$ distinct operations required by a protocol can be mapped to unique register states. This provides insights into how to choose $r$ that balances programmability against protocol latency in a slotted controller-driven network.

\subsection{Linear Combination of Unitaries (LCU) and Kraus Formulation}\label{ssec:kraus}

Equation~\eqref{eq:Kraus-operator} can be recast as a general operator acting on the electron spin subsystem:
\begin{equation}
K_{\phi,\psi_N} = 
\sum_{k=0}^{K-1} d_k^{\!*} c_k\, U_k,
\end{equation}. 

Upon obtaining outcome $\ket{\phi}$, the electron spin state undergoes the transformation
\begin{equation}
\ket{\psi_{E_0}} 
\longmapsto 
\frac{K_{\phi,\psi_N}\ket{\psi_{E_0}}}
{\big\|K_{\phi,\psi_N}\ket{\psi_{E_0}}\big\|},\\
p_\phi 
= \big\|K_{\phi,\psi_N}\ket{\psi_{E_0}}\big\|^2,
\end{equation}
where $p_\phi$ is the probability of the outcome.

For a general mixed input state of electron spin $\rho_E$, this becomes $p_\phi = \mathrm{Tr}[K_{\phi,\psi_N}\rho_E K_{\phi,\psi_N}^\dagger]$. 
More generally, an orthonormal measurement $\{\ket{\phi_j}\}_j$ on the nuclear register, with $\ket{\phi_j} = \sum_k d_{j,k}\ket{k}$, defines a \emph{quantum instrument} on the electron spin  (Eq.~\eqref{eq:Kraus-operator}):
\begin{equation}
\begin{gathered}
K_j = \bra{\phi_j} U_{\mathrm{repeater}} \ket{\psi_N}
    = \sum_k d_{j,k}^{\!*} c_k\, U_k, \\
\mathcal{E}_j(\rho_E) = K_j \rho_E K_j^\dagger,
\end{gathered}
\label{eq:kraus_operator_final}
\end{equation}

where each $\mathcal{E}_j$ is a completely positive (CP), trace-non-increasing map with probability $p_j = \mathrm{Tr}[K_j\rho_E K_j^\dagger]$. 
The overall transformation $\sum_j \mathcal{E}_j$ is completely positive and trace preserving (CPTP).

Equation~\eqref{eq:kraus_operator_final} directly implements a \emph{Linear Combination of Unitaries (LCU)}:
\begin{equation}
K_j = \sum_k \alpha_{j,k} U_k, 
\qquad \alpha_{j,k} = d_{j,k}^{\!*} c_k,
\end{equation}
linking the nuclear-register superposition coefficients to the effective combination weights. This formalism connects node-level programmability to established techniques in Hamiltonian simulation and quantum channel decomposition~\cite{childshamiltonian12,chakrabortylinearcombination24}.

%A related mathematical structure appears in the recent ``quantum path'' framework, or \emph{quantum SWITCH}, proposed in \cite{caleffiquantumswitch23} that places the order of two channels in coherent superposition. Our ISA instead uses a local nuclear register to coherently select among electron spin-side unitaries within a node. Extending the ISA to noisy CPTP operations and order superposition is interesting future work.

\section{Conclusion}\label{sec:conclusion}

We introduced programmable instruction-set architecture (ISA)-style abstraction for NV-center quantum repeater nodes, enabling controller-driven programmability at the hardware level. Two control modes were introduced: deterministic register control, analogous to classical flow-rule execution, and coherent register control, which allows quantum superpositions of operations. The coherent mode yields a heralded Kraus or LCU description that enables interference-based diagnostics such as fidelity witnessing and in situ calibration. Using the BBPSSW purification protocol as an example, we illustrated how network-level coordination can be expressed as controller-issued instruction sequences. The framework extends naturally to multi-spin and multi-node settings and establishes a pathway toward flexible, scalable, and diagnostically aware quantum networking architectures.

%\section*{Acknowledgment}

\clearpage
\appendices

\section{Instruction format for BBPSSW purification protocol example}
The instruction format to execute the BBPSSW purification protocol in section~\eqref{ssec:controllerdrivenexample} is as follows:
\begin{enumerate}
    \item \textbf{Bilateral twirling to form rotationally symmetric Werner states.}
    The controller issues
    \[
    \mathbf{I}(t) = \Big(\INSTR^{(A,e)}(t),\,\INSTR^{(B,e)}(t)\Big),
    \]
\[
\begin{aligned}
\Rightarrow\;
\mathbf{I}(t)
&= \Big(
   \big(\INSTR^{(A, 0)}(t),\, \INSTR^{(A, 1)}(t)\big), \\
&\qquad
   \big(\INSTR^{(B, 0)}(t),\, \INSTR^{(B, 1)}(t)\big)
  \Big).
\end{aligned}
\]

    where each instruction has
\begin{equation}
\begin{aligned}
\textsf{OPCODE} &= \textsf{rand}\{I, X, Y, Z\} = u (say), \\
\textsf{PARAMS} &= \emptyset, \\
\textsf{PATTERN}   &= \textsf{PATTERN}(u), \\
\textsf{MODE}   &= \textsf{deterministic}.
\end{aligned}
\label{eq:pauli_fields}
\end{equation}

    Both nodes apply the same random single-qubit rotation $u$ to their qubits, effectively depolarizing each pair into a rotationally symmetric Werner state.

    \item \textbf{Bilateral CNOT from “pair to keep” to “pair to measure”.}
    The controller issues
    \[
\begin{aligned}
\INSTR^{(A)}(t{+}1)
&= \big(
   \textsf{CNOT},\;
   \textsf{PARAMS}=(E_{A1}\!\to\!E_{A2}), \\
&\qquad
   100,\;
   \textsf{deterministic}
  \big),
\end{aligned}
\]

and similarly for node $B$ as $E_{B1}\!\to\!E_{B2}$
\[
\begin{aligned}
\INSTR^{(B)}(t{+}1)
&= \big(
   \textsf{CNOT},\;
   \textsf{PARAMS}=(E_{B1}\!\to\!E_{B2}), \\
&\qquad
   100,\;
   \textsf{deterministic}
  \big),
\end{aligned}
\]
so that both nodes perform a CNOT with the keep-pair qubit as control and the measure-pair qubit as target.

    \item \textbf{Local measurement of target qubits.}
    The controller issues
\[
\begin{aligned}
\INSTR^{(A)}(t{+}2)
&= \big(
   \textsf{MEASURE},\;
   \textsf{PARAMS}=(E_{A2}, \\
&\qquad\textsf{basis}=Z),\;
   101,\;
   \textsf{deterministic}
  \big),
\end{aligned}
\]

    and similarly for node $B$ on $E_{B2}$
    \[
\begin{aligned}
\INSTR^{(B)}(t{+}2)
&= \big(
   \textsf{MEASURE},\;
   \textsf{PARAMS}=(E_{B2}, \\
&\qquad\textsf{basis}=Z),\;
   101,\;
   \textsf{deterministic}
  \big).
\end{aligned}
\] Each node measures its target qubit in the computational basis and sends the classical outcome to the other node.

    \item \textbf{Parity check and sifting.}
    If the measurement outcomes are identical (even parity), the control pair $(E_{A1}, E_{B1})$ is kept; otherwise, it is discarded. This sequence is iterated over surviving pairs to progressively raise their fidelity.
\end{enumerate}

\section{Entanglement transfer from electron-electron spin to nuclear-nuclear spin}
\begin{figure}[tb]
    \centering
    \includegraphics[width=\columnwidth]{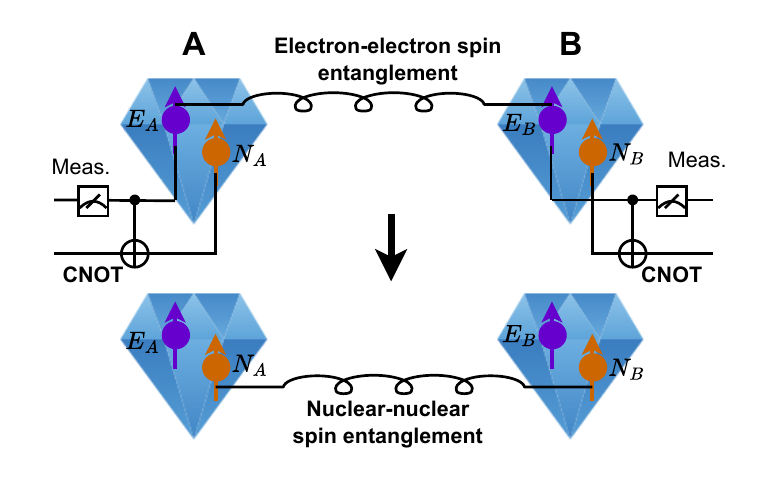}
    \caption{Entanglement transfer from electron-electron spin to nuclear-nuclear spin in neighbouring NV-center nodes.}
    \label{fig:NV_entanglement_transfer}
    %\vspace{-0.5cm}
\end{figure}
The entangling operation in section~\eqref{ssec:deterministiccontrol} corresponding to selection of $N_1 = 1, \ N_2 = 1$ in deterministic mode transfers the entanglement from initial electron-electron spin to long-lived nuclear-nuclear spin (ancillary nuclear spin at each node involved).
Figure~\ref{fig:NV_entanglement_transfer} schematically illustrates the corresponding procedure in a network setting, in which entanglement initially established between electron spins is coherently transferred to nuclear spins. The protocol proceeds as follows.

Consider two neighbouring NV centres, denoted by nodes A and B, whose electron spins are prepared in a maximally entangled Bell state as shown in Fig.~\ref{fig:NV_entanglement_transfer}:
\begin{equation}
    \ket{\Psi_{E_A E_B}} = \frac{1}{\sqrt{2}}\bigl( \ket{0}\ket{0} + \ket{1}\ket{1} \bigr)_{E_A E_B}.
\end{equation}
Assuming that the nuclear spins at both nodes are initialised in the state $\ket{0}$, the combined electron–nuclear system is described by
\begin{equation}
    \ket{\Psi_{\text{initial}}}
    = \frac{1}{\sqrt{2}}\bigl( \ket{0}_{E_A}\ket{0}_{E_B} + \ket{1}_{E_A}\ket{1}_{E_B} \bigr)
    \otimes \ket{0}_{N_A}\ket{0}_{N_B}.
\end{equation}

Next, local controlled-NOT (CNOT) gates are applied independently at both nodes, using the electron spins as control qubits and the nuclear spins as target qubits, i.e.,
\begin{equation}
    \mathrm{CNOT}_{E \rightarrow N} \qquad \text{(applied at nodes A and B)}.
\end{equation}
As a result, the joint state of the system evolves to
\begin{equation}
\begin{aligned}
\ket{\Psi_{\text{after CNOT}}}
= \frac{1}{\sqrt{2}}\bigl(
&\ket{0}_{E_A}\ket{0}_{N_A} \otimes \ket{0}_{E_B}\ket{0}_{N_B} \\
&+ \ket{1}_{E_A}\ket{1}_{N_A} \otimes \ket{1}_{E_B}\ket{1}_{N_B}
\bigr).
\end{aligned}
\end{equation}

Finally, a projective measurement\footnote{Even in the absence of explicit measurement on the electron spins, the nuclear spins remain entangled. The measurement is included here solely for clarity of presentation.} is performed on the electron spins. Conditioning on the measurement outcome, the nuclear spins are left in an entangled Bell state:
\begin{equation}
    \ket{\Psi_{N_A N_B}} = \frac{1}{\sqrt{2}}\bigl( \ket{0}_{N_A}\ket{0}_{N_B} + \ket{1}_{N_A}\ket{1}_{N_B} \bigr).
\end{equation}

In the discussion above in section~\eqref{ssec:deterministiccontrol}, this transfer process is emulated locally by applying an entangling gate between the electron spin and a nuclear spin under the control condition $N_1 = 1$ and $N_2 = 1$.

\section{Derivation of probabilities equation for two-branch interference example}
After Eq.~\eqref{eq:two-branch-after} in section~\eqref{ssec:coherentcontrol} we measure the nuclear spin in the $X$ basis,
\(
\ket{\phi_\pm}=(\ket{0}\pm\ket{1})/\sqrt{2},
\)
yields the unnormalized electron states

\vspace{-0.5em}
\par\noindent\rule{\linewidth}{0.4pt}
\vspace{-1.2em}
\begin{strip}
\vspace{-1.5em}
\begin{equation}
(\bra{\phi_\pm}\otimes I)\ket{\Psi_{\mathrm{after}}}
= \frac{1}{\sqrt{2}}
\Big(\tfrac{1}{\sqrt{2}}\bra{0}\pm\tfrac{1}{\sqrt{2}}\bra{1}\Big)
\Big(\ket{0}\otimes U_0\ket{\psi_{E_0}}
+ \ket{1}\otimes U_1\ket{\psi_{E_0}}\Big).
\end{equation}
\vspace{-1.5em}
\end{strip}
\vspace{-1.2em}
\noindent\rule{\linewidth}{0.4pt}
\vspace{-0.5em}

\begin{equation}
\Rightarrow (\bra{\phi_\pm}\otimes I)\ket{\Psi_{\mathrm{after}}}
= \tfrac{1}{2}\Big(U_0\ket{\psi_{E_0}}\pm U_1\ket{\psi_{E_0}}\Big),
\end{equation}
\begin{align}
 \Rightarrow \ket{\psi_{E_0}^{(\pm)}}&= (\bra{\phi_\pm}\otimes I)\ket{\Psi_{\mathrm{after}}} = \frac{1}{2}\Big(U_0 \;\pm\; U_1\Big) \ket{\psi_{E_0}}.
\end{align}
Hence, the corresponding success probability is:

\begin{align}
p_\pm
&= \left\langle \psi_{E_0} \left| \tfrac{1}{2}(U_0^\dagger \pm U_1^\dagger)\, 
                      \tfrac{1}{2}(U_0 \pm U_1) \right| \psi_{E_0} \right\rangle, \notag \\
p_\pm
&= \tfrac{1}{4} \Big( \langle \psi_{E_0} | U_0^\dagger U_0 | \psi_{E_0} \rangle
   + \langle \psi_{E_0} | U_1^\dagger U_1 | \psi_{E_0} \rangle \notag \\
&\quad \pm \langle \psi_{E_0} | U_0^\dagger U_1 | \psi_{E_0} \rangle
   \pm \langle \psi_{E_0} | U_1^\dagger U_0 | \psi_{E_0} \rangle \Big). \label{eq:p_pm_expanded}
\end{align}

Since $U^{\dagger} U =I$ so $\langle \psi_{E_0} | U_0^{\dagger} U_0 | \psi_{E_0} \rangle = \langle \psi_{E_0} | U_1^{\dagger} U_1 | \psi_{E_0} \rangle = \langle \psi_{E_0} | I | \psi_{E_0} \rangle = 1$. 

And since $\langle \psi_{E_0} | U_0^\dagger U_1 | \psi_{E_0} \rangle$ is hermitian conjugate of $\langle \psi_{E_0} | U_1^\dagger U_0 | \psi_{E_0} \rangle$ i.e., $\langle \psi_{E_0} | U_0^\dagger U_1 | \psi_{E_0} \rangle$ = $\langle \psi_{E_0} | U_1^\dagger U_0 | \psi_{E_0} \rangle^{\dagger}$. Hence $\langle \psi_{E_0} | U_0^{\dagger} U_1 | \psi_{E_0} \rangle + \langle \psi_{E_0} | U_1^{\dagger} U_0 | \psi_{E_0} \rangle = 2 \ \mathrm{Re} \langle \psi_{E_0} | U_0^{\dagger} U_1 | \psi_{E_0} \rangle$ which gives
\begin{align}
\Rightarrow p_\pm &= \tfrac{1}{4} \Big( 1 + 1 \pm 2\,\mathrm{Re}\,\langle \psi_{E_0} | U_0^\dagger U_1 | \psi_{E_0} \rangle \Big),
\end{align}

\begin{equation}\label{eq:p_pm_basic}
\Rightarrow p_\pm = \tfrac{1}{2}\big(1 \pm \mathrm{Re}\langle\psi_{E_0}|U_0^\dagger U_1|\psi_{E_0}\rangle\big).
\end{equation}

And the corresponding (normalized) electron states after measurement of nuclear qubit is as follows:
\begin{equation}\label{eq:electronpm_map}
\ket{\psi_{E_0}^{(\pm)}} = \frac{\big(U_0 \pm U_1\big)\ket{\psi_{E_0}}}{\sqrt{\langle \psi_{E_0} | (U_0^{\dagger} \pm U_1^{\dagger}) \big(U_0 \pm U_1\big) | \psi_{E_0} \rangle}},
\end{equation}
\begin{equation}\label{eq:estate_overlap_nophase}
\Rightarrow \ket{\psi_{E_0}^{(\pm)}} \;\propto\; \big(U_0 \pm U_1\big)\ket{\psi_{E_0}}.
\end{equation}
From Eq.~\eqref{eq:p_pm_basic} and Eq.~\eqref{eq:estate_overlap_nophase}, we know the information about the interference term (the mapped state after the measurement and its probability) from the application of the two branches of unitary.

\section{Extracting out the complete state with two-phase settings in coherent mode}
 Using $e^{i\varphi} = \cos{\varphi}+i \sin{\varphi}$ in Eq~\eqref{eq:p_pm_phi_final} we have:

\begin{equation}
\begin{aligned}
\Rightarrow\;
p_{\pm}(\varphi)
&= \tfrac{1}{2}\Big(1 \ \pm \mathrm{Re}\Big[
     \cos{\varphi}\,\bra{\psi_{E_0}} U_0^{\dagger} U_1 \ket{\psi_{E_0}} \\
&\qquad\quad + i \sin{\varphi}\,\bra{\psi_{E_0}} U_0^{\dagger} U_1 \ket{\psi_{E_0}}
   \Big]\Big).
\end{aligned}
\label{eq:p_pm_phi_reim}
\end{equation}

Taking just $p_+$ measurements with $\varphi_1=0$ and $\varphi_2=-\pi/2$. Let $a = \bra{\psi_{E_0}} U_0^{\dagger} U_1 \ket{\psi_{E_0}}$.
\begin{equation}
\begin{aligned}
\text{For} \ \varphi_1 &= 0 \\
\Rightarrow\quad
p_+^{(1)} &= \tfrac12\big(1+\mathrm{Re}[\,1\cdot a + i\cdot 0\cdot a\,]\big) \\
&= \tfrac12\big(1+\mathrm{Re}\,a\big).
\end{aligned}
\label{eq:phi1_case}
\end{equation}

\begin{equation}
\begin{aligned}
\text{For} \ \varphi_2 &= -\tfrac{\pi}{2} \\
\Rightarrow\quad
p_+^{(2)} &= \tfrac12\big(1+\mathrm{Re}[\,0\cdot a + i\cdot(-1)\cdot a\,]\big) \\
&= \tfrac12\big(1+\mathrm{Im}\,a\big).
\end{aligned}
\label{eq:phi2_case}
\end{equation}

Inverting these to recover the overlap amplitude:
\begin{equation}
\mathrm{Re}\,a \;=\; 2\,p_+^{(1)} - 1,
\qquad
\mathrm{Im}\,a \;=\; 2\,p_+^{(2)} - 1.
\end{equation}
Hence, the state is:
\begin{equation}\label{eq:amplitude_calculated}
    \Rightarrow a = \bra{\psi_E} U_0^{\dagger} U_1 \ket{\psi_E} = (2 p_+^{(1)} - 1) + i\, (2 p_+^{(2)} - 1).
\end{equation}
\end{document}